\documentclass[sigconf, nonacm]{acmart}

\usepackage{algorithm}
\usepackage{algpseudocode}
\usepackage{subfigure}
\usepackage{makecell}

\usepackage{comment}

\algnewcommand\And{\textbf{and}}

\newcommand{\cut}[1]{}

\usepackage{tensor}
\newcommand*{\Perm}[2]{{}^{#1}\!P_{#2}}%

\begin{document}
\title{Identifying Possible Winners in Ranked Choice Voting Elections with Outstanding Ballots}

\author{Alborz Jelvani}
\affiliation{%
  \institution{Rutgers University}
}
\email{aj654@scarletmail.rutgers.edu}

\author{Amélie Marian}
\affiliation{%
  \institution{Rutgers University}
}
\email{amelie.marian@rutgers.edu}

\begin{abstract}
 Several election districts in the US have recently moved to ranked-choice voting (RCV) to decide the results of local elections. RCV allows voters to rank their choices, and the results are computed in rounds, eliminating one candidate at a time. RCV ensures fairer elections and has been shown to increase elected representation of women and people of color. 
A main drawback of RCV is that the round-by-round process requires all the ballots to be tallied before the results of an election can be calculated. With increasingly large portions of ballots coming from absentee voters, RCV election outcomes are not always apparent on election night, and can take several weeks to be published, leading to a loss of trust in the electoral process from the public.
In this paper, we present an algorithm for efficiently computing possible winners of RCV elections from partially known ballots and evaluate it on data from the recent New York City Primary elections. We show that our techniques allow to significantly narrow down the field of possible election winners, and in some case identify the winner as soon as election night despite a number of yet-unaccounted absentee ballots, providing more transparency in the electoral process.

\end{abstract}

\maketitle

\section{Introduction}
Ranked-choice voting (RCV) --also commonly referred to as Instant-Runoff Voting (IRV), or Single Transferable Vote (STV)-- is a voting mechanism that allows voters to rank candidates in their order of preference. Counting typically proceeds in rounds, by eliminating the candidate with the lowest number of votes and transferring that candidate's votes to the next candidate on each of the voters' preference lists. The process continues until the leading candidate receives more than 50\% of active votes. 

Ranked Choice voting is used in national elections in several countries, such as Australia, Ireland, and the United Kingdom.
In the U.S. multiple counties and municipalities  have recently adopted RCV for their elections~\cite{fairvote} with positive impacts in voter participation~\cite{kimball2016voter,mcginn2020rating,Juelich2021RankedCV}. 
Ranked Choice Voting has gained traction because of several advantages: it is seen as a fairer way to run elections~\cite{mill1859considerations} and to improve  representation of women~\cite{terrel2021,JOHN201890}, voters of color~\cite{FairVoteVOC,JOHN201890} and participation in youth voters~\cite{Juelich2021RankedCV}. It avoids the “spoiler effect,” reduces ``wasted" votes when many candidates are running, and saves money by avoiding runoff elections. In addition, by aiming at forming a consensus behind the selected candidates, RCV decreases  incentives for strategic voting. 

However, moving to RCV  has some disadvantages: voters may be confused by the ranking mechanism and the election results can take a long time to be processed as the vote transfers are only tabulated once all votes are in.
This limitation is due to the round-by-round vote counting process of RCV. To move to the next round, {\em all} votes need to be tallied to accurately eliminate the candidate with the lowest number of votes. In query processing terms, each round is a non-monotonic blocking operation~\cite{blocking}, and the process can only continue once the round is fully tabulated. This is often cited as one of the main drawbacks of RCV\cite{atlantic2019} as it delays the results of the election until all votes are gathered, which can take several days or weeks in counties with a large number of mail-in or absentee votes. In 2018 the San Francisco mayoral results took a week to be tabulated and confirmed in large part due to the late counting of mail-in ballots. In NYC, the June 2021 primary results were certified {\em a full month} after the election due to the large number of absentee ballots; preliminary RCV results were not made available to the public for over a week after the election, and did not originally include absentee ballots. These delays, the lack of transparency, and the incomplete information, or lack thereof, on the outcome of cast ballots lead to distrust in the RCV election process from a population that is used to having election results, or at least close estimates, soon after an election. 

In this paper, we present an algorithm to process partial results of RCV races  without requiring all votes to be gathered before the counting can start (Section~\ref{sec:algo}). Our novel approach considers voter preferences from tallied ballots to identify possible elimination orderings based on the voting data that is already known, and taking into account the uncertainty associated with still-missing (e.g., absentee) ballots (Section~\ref{sec:verify}). We propose a branch and bound algorithm to speed up the search (Section~\ref{sec:branchbound}).
Our algorithm would allow for identifying candidates who still have a path to victory, and those who do not, as soon as election night, providing stakeholders with more transparency on the election outcome. 

We evaluated our algorithm on election night data from the NYC 2021 Primary elections in Section~\ref{sec:mainresults}, and were able to identify,  with certainty, only one possible winner in about 40\% of the races with more than two candidates, even with absentee ballots still outstanding.
 We report on several of these races in detail in Section~\ref{sec:casestudies}.


\section{Background and Related Work}

Ranked Choice Voting generally describes any electoral system that allows voters to list candidates in their order of preference. These preferences may be aggregated using several different vote counting algorithms~\cite{sep-voting-methods}. In the U.S., the method of counting typically involves a series of rounds where candidates are eliminated  and where votes for eliminated candidates are transferred to next preferred candidates on the voters' lists. The names RCV, IRV (Instant-Runoff Voting), and SVT (Single Vote Transfer) have been used interchangeably to describe similar electoral systems around the world: RCV elections are being used in several countries such as Australia, Canada, the U.K., and New Zealand.

We focus on the impact of yet-unaccounted ballots on RCV election results, a scenario prompted by election rules in the U.S. where a portion of the votes can be cast through absentee ballots. The presence of absentee ballots can significantly delay election results; several states allow for absentee ballots to be postmarked until the day of the election, and received up to 17 days {\em after the election} (California)\footnote{\url{https://www.vote.org/absentee-ballot-deadlines/}}. New York requires a waiting period of at least a week before processing absentee ballots~\cite{NYSAbsentee}. After many states relaxed their rules due to the Covid-19 pandemic, the number of mail-in ballots increased significantly: 46\% of ballots in the November 2020 election were mail-in ballots~\cite{electionsurveyUS2021}. While the results of traditional majority- or plurality-based elections can easily be estimated even with a large proportion of outstanding ballots (with appropriate confidence intervals), the results of RCV elections cannot accurately be processed until all ballots are cast, as new ballots may result in a different elimination order, which in turn may result in a different allocation of vote transfers and a significantly different outcome. The arrival of ballots can thus be seen as a data stream of ballots with unpredictable (but slow) arrival rates~\cite{datastream}, and the vote counting in a RCV election as a non-monotonic blocking operation~\cite{blocking}, on which traditional online aggregation methods~\cite{onlineagg} or adaptive query processing techniques~\cite{adaptive} for unpredictable data arrival are not applicable as they rely on pipelined operations. Several localities have opted to provide a temporary RCV count of partial ballot data as soon as election night, but partial results of non-monotonic operators may be incorrect~\cite{partial}; in a RCV election scenario this may lead to loss of trust in the electoral process if the results change significantly due to new data from outstanding ballots.



We propose an algorithm to identify all still-possible election outcomes, considering the already-known votes, including their preference orderings, and the uncertainty associated with ballots that still need to be tallied. This is related to querying possible worlds~\cite{possibleworlds} on an incomplete dataset~\cite{incompleteDB}, where the partially known ballot information is the incomplete dataset, the RCV vote counting is a the query, and the set of possible worlds are all possible elimination paths and the resulting winners. Our techniques explore this set of possible worlds, narrows down candidates to a set of still-possible winners, and shows the elimination paths that lead to each winning outcome, along with an estimation of the minimum number of unknown ballots a candidate would need to win.

The use of RCV in real-world elections has led to recent work that studies the intersection of voting theory with regulatory frameworks. In particular, there has been an interest in defining and computing the margin of victory for RCV elections in Australia, where small margins would trigger elections audits~\cite{blom2016efficient,magrino2011computing}.
Our work is related to this work, to studies in the shift in the balance of power~\cite{blom2020power,blom2020lostballot}, and to work on election manipulation~\cite{blom2019election} in RCV settings. However, unlike these works, we do not aim to identify the minimum number of vote changes that would trigger a permutation in the elimination order, but focus on finding all possible elimination orders given a number of unknown (unbound) ballots.

\section{Definitions}

\paragraph{Candidates}

An election is performed over $\mathcal{C} = \{c_1,c_2, ..., c_n\}$, the set of $n$ candidates running in the election.

\paragraph{Ballots}

Ballots are votes cast by individual voters. For each election, a voter casts exactly one ballot.

\begin{itemize}

\item A \textbf{ballot signature} is defined as an r-tuple $(c_1, c_2, ..., c_r)$ for  $c_i \in C$, where $r \leq min(n_c,n)$, where $n_c$ is the maximum number of choices allowed in the election\footnote{Rules vary: some elections in Australia requires voters to rank all candidates (full ranking), San Francisco allows up to 10 choices, New York only allows up to five.} and candidates are listed in order of preference, with $c_1$ as the most preferred candidate. A ballot with  $r < n$ is a \textit{partial ranking} of the candidates. 

\item A \textbf{bound ballot} is a ballot for which the r-tuple $(c_1, c_2, ..., c_r)$ is known. For example, on election night, all in-person ballots are typically bound. 

    
    
    \item An \textbf{unbound ballot} is a valid ballot for which the r-tuple $(c_1, c_2, ..., c_r)$ is not known yet. This is typically the case of mail-in ballots that have not been opened yet, or any ballot for which the validity status is contested.

\end{itemize}

Ballots in a RCV election are comprised of
$\mathcal{B}$, the set of bound ballots, and $\mathcal{U}$, the set of unbound ballots. The election cannot be certified until all ballots in $\mathcal{U}$ have been moved to $\mathcal{B}$ or voided. The final election results are computed over $\mathcal{B}$ once $\mathcal{U}=\emptyset$

\paragraph{Elections} An election is a voting process over a set of candidates and ballots.

\begin{itemize}

    \item An \textbf{election profile} is defined by the 3-tuple $(\mathcal{C},\mathcal{B},\mathcal{U})$ where $\mathcal{C} = \{c_1,c_2, ..., c_n\}$ is the set of $n$ candidates running in the election. $\mathcal{B}$ is the set of bound ballots, and $\mathcal{U}$ is the set of unbound ballots. We define $m = |\mathcal{B}| + |\mathcal{U}|$ as the total number of ballots in an election profile.
    
    \item An \textbf{elimination order}, $\pi$, is a permutation of $\mathcal{C}$. It is represented as an ordered tuple $(e_1,e_2,...,e_n)$ where $e_1$ is the first candidate to be eliminated and $e_n$ represents the winner of the elimination order.
    
    \item The \textbf{search-space}, $\mathcal{S}$, of an election profile is the collection of all possible elimination orders.
    
\end{itemize}

\paragraph{Vote counting.} The following definitions are to assist in the vote counting process. 

\begin{itemize}

    \item A \textbf{partial elimination order}, $\pi'$, is a prefix of a permutation order $\pi$ such that candidates $(e_1,e_2,...,e_j)$, $j \leq n$ are  eliminated.

   \item The \textbf{set of still-active candidates $\mathcal{C'}$} is the set of candidates $\mathcal{C} \setminus \pi'$ who have not been yet eliminated after applying a partial elimination order $\pi'$.

    \item The \textbf{tally count} of a candidate $c \in \mathcal{C'}$, denoted as $t_c$, is defined as the current number of ballots $b \in \mathcal{B}$ such that $c$ is the highest ranked candidate $c_i \in \mathcal{C'}$ in $b$'s r-tuple.

    \item A ballot $b$ is considered \textbf{exhausted} when  all candidates $c_i$ of $b$'s r-tuple $(c_1, c_2, ..., c_r)$ have been eliminated  ($c_i \not \in \mathcal{C'}$). 
    
    
 \end{itemize}

\section{Tabulating partial RCV results}
\label{sec:algo}

The general algorithm for counting votes in a RCV election~\cite{nanson_1864}, only applies when all ballots are known and bound ($\mathcal{U} = \emptyset$). In elections where some ballots are outstanding, such as absentee ballots or ballots in dispute, running the general vote counting algorithm on partial results runs the risk of returning misleading information as the permutation order $\pi_{partial}$ may turn out to be very different from the actual $\pi$ elimination order on the full set of election ballots. A small change in the relative ordering of two (even minor) candidates can cause a ripple effect that changes the outcome of the election. Many municipalities have chosen to make public the results of this general vote counting process on the ballots known on election night, sometimes resulting in confusing information for the public as the final results may differ.

We address this problem by proposing an algorithm (Algorithm~\ref{alg:rcv_branch}) that considers all possible elimination orders that may still be possible under the constraints given by known bound ballots in $\mathcal{B}$. Our approach considers unknown, unbound ballots in $\mathcal{U}$ and identifies all possible configurations of outcomes if (subsets of) these ballots were bound to each candidate in $\mathcal{C}$. To process our algorithm, we need to make tentative (or mock) assignments for unbound ballots to test possible alternatives. We thus define:
\begin{itemize}
    \item The set of tentatively bound ballots $\mathcal{B'}$ which contains the set of bound ballots $\mathcal{B}$ and a set of unbound ballots from $\mathcal{U}$ for which we have made tentative bindings by hypothetically assigning them to a (set of candidates). Initially, $\mathcal{B'}$=$\mathcal{B}$.
    \item The set of tentatively unbound ballots $\mathcal{U'}$ which contains the set of unbound ballots $\mathcal{U}$ minus those ballots for which we have made tentative bindings by hypothetically assigning them to a (set of candidates) and that were moved to $\mathcal{B'}$. Initially, $\mathcal{U'}$=$\mathcal{U}$.
\end{itemize}

Ballots are moved from $\mathcal{U'}$ to $\mathcal{B'}$ when a tentative (hypothetical) assignment is made for a candidate to make a specific elimination order possible. Ballots can be moved from $\mathcal{B'}$ to $\mathcal{U'}$ if they were originally in $\mathcal{U}$, all the candidates on the ballot have been eliminated, and the ballot is not exhausted.

\subsection{Verifying the Possibility of an Election Outcome}
\label{sec:verify}

The main challenge in evaluating partial RCV election results is that the space of possible outcomes is exponential in the number of unknown ballots. Evaluating each ballot permutation is impossible. 
However, if we only want to identify whether a candidate has a path to victory, we only need to selectively explore the space of all elimination orders, and verify if each path in the elimination order tree is possible with the currently known, bound, ballots and some permutation of the unknown, unbound, ballots.  

The main contribution of this paper is the \texttt{verify} function (Algorithm~\ref{alg:rcv_verify}), which takes an (possibly partial) elimination order $\pi'$ as input and given a tentative election profile ($\mathcal{C'},\mathcal{B'},\mathcal{U'}$) consisting of a set of remaining candidates $\mathcal{C'}$, and sets of bound ballots $\mathcal{B'}$ and unbound ballots $\mathcal{U'}$ returns \texttt{True} if the elimination order is possible.

\begin{algorithm}[]
\caption{Function to check if the (partial) elimination order $\pi'$ is possible under an election profile with tentative candidates $\mathcal{C'}$, tentative bound ballots $\mathcal{B'}$, and tentative unbound ballots $\mathcal{U'}$.}
\label{alg:rcv_verify}
    \begin{algorithmic}[1]
    \Function{verify}{$\pi'$,$(\mathcal{C'},\mathcal{B'},\mathcal{U'})$}
        \For{each $e \in \pi'$}
            \For{each $b \in \mathcal{B'}$} \Comment{Loop 1: Tally bound ballots}
                \If{there exists a highest ranked $c \in b$ such that $c \in \mathcal{C'}$}
                    \State $t_c(c) \gets t_c(c) + 1$
                \Else
                    \If{$b$ is marked as an absentee ballot and $b$ is not exhausted}
                        \State $\mathcal{B'} \gets \mathcal{B'} \setminus \{b\}$
                        \State $\mathcal{U'} \gets \mathcal{U'} \cup \{b\}$
                    \EndIf
                \EndIf
            \EndFor

            \For{each $c \in \mathcal{C'} \setminus \{e\}$ such that $t_c(c) \leq t_c(e)$} \Comment{Loop 2: Assign unbound ballot to force elimination order}
                \State $margin \gets t_c(e) - t_c(c) + 1$ \Comment{the number of votes needed to eliminate the next candidate in $\pi'$}
                \If{$|\mathcal{U'}|- margin < 0$}
                    \State \Return $False$ \Comment{Not enough unbound votes to force elimination order $\pi'$}
                \EndIf
                \While{$margin \neq 0$}
                    \State Assign $c$ to next ranking of some $b' \in \mathcal{U'}$
                    \State $\mathcal{U'} \gets \mathcal{U'} \setminus \{b'\}$
                    \State $\mathcal{B'} \gets \mathcal{B'} \cup \{b'\}$
                    \State $margin \gets margin -1$
                \EndWhile
                
            \EndFor

            \State $\mathcal{C'} \gets \mathcal{C'} \setminus \{e\}$
            \State Reset all tally counts to 0
        \EndFor
        
    \State \Return $True$
    \EndFunction
    \end{algorithmic}
\end{algorithm}

Function \texttt{verify} first tallies the number of bound votes (from ballots in $\mathcal{B'}$) that each candidate $c \in \mathcal{C'}$ receives (Loop 1). Then if $e$, the next candidate to be eliminated in $\pi'$, is not the candidate with the lowest number of votes, \texttt{verify} binds some ballots in  $\mathcal{U'}$ to the candidate(s) with fewer votes than $e$ and moves them to $\mathcal{B'}$, to force the elimination of $e$ (Loop 2). If there are not enough remaining ballots in $\mathcal{U'}$, the elimination order is considered impossible and \texttt{verify} returns false.

To achieve this, at each round \texttt{verify} attempts to boost the vote count of each candidates ranked below $e$ to pass $e$ by only 1 vote. This ensures that each boosted candidate outranks $e$ by using the minimum amount of unbound ballots. Thus, if there are not enough unbound ballots to boost a candidate past $e$ then the corresponding elimination order is indeed infeasible as all previous boosts used the minimum possible amount of unbound ballots to arrive at the current elimination round.

The runtime of \texttt{verify} is $O(|\mathcal{C}|^2m)$.
The outer loop iterates at most $|\mathcal{C}|$ times and Loops 1 and 2 each iterate ($|\mathcal{B}| + |\mathcal{U}|$)=$m$ and $|\mathcal{C}|$ times respectively. The while loop inside Loop 2 also iterates $O(m)$ times. Therefore our total runtime is given by $O(|\mathcal{C}|(m + m|\mathcal{C}|)) = O(|\mathcal{C}|^2m)$. 

Note that \texttt{verify} considers all possible worlds, including those where one candidate would receive {\em all} outstanding votes. An interesting avenue to explore in future work would be to cap the number of votes candidates can receive, possibly based on poll projections, to estimate the chances of each candidate and provide better approximate predictions, with some probabilistic bounds guarantees.

\subsection{Calculating all Possible Election Outcomes}
\label{sec:branchbound}

A brute-force approach to identify all possible outcomes of an election given a set of bound ballots $\mathcal{B}$ and a set of unbound ballots $\mathcal{U}$ would consider the complete search space $\mathcal{S}$ of all permutations $\pi$ of candidates in $\mathcal{C}$ and call \texttt{verify($\pi,(\mathcal{C},\mathcal{B},\mathcal{U})$)} for each one. This brute force approach will explore the entire search-space, but requires $|\mathcal{C}|!$ iterations. Thus, a lower bound on its runtime is given by $\Omega(|\mathcal{C}|!)$, a prohibitively expensive approach.

To circumvent the prohibitive cost of the brute-force approach, we introduce \texttt{calculatePossibleOutcomes} (Algorithm~\ref{alg:rcv_branch}), a branch and bound algorithm to identify all possible elimination paths given the known ballots $\mathcal{B}$ and unbound ballots $\mathcal{U}$. Our algorithm selectively prunes elimination orders (and elimination order prefixes) when these cannot be verified (Algorithm~\ref{alg:rcv_verify}) given the current election profile. The algorithm keeps track of the ballots that have been tentatively bound ($\mathcal{B'}$), and the corresponding remaining unbound ballots ($\mathcal{U'}$) to reach each partial elimination order prefix $\pi'$. The \texttt{calculatePossibleOutcomes} function is recursive; it is initiated from an external routine by calling \texttt{calculatePossibleOutcomes($\emptyset,(\mathcal{C},\mathcal{B},\mathcal{U})$)}.

\begin{algorithm}[h]
\caption{Recursive algorithm that enumerates all possible elimination orders $\mathcal{S}$ for an election profile $(\mathcal{C},\mathcal{B},\mathcal{U})$.  }  
\label{alg:rcv_branch}
    \begin{algorithmic}[1]
    \State $\mathcal{S} \gets \emptyset$
    \State $\pi' \gets \emptyset$
    \Procedure{calculatePossibleOutcomes}{$\pi'$,$(\mathcal{C},\mathcal{B},\mathcal{U})$}
        \If{$|\pi'| == n$}
            \State Add $\pi'$ to $\mathcal{S}$
            \State \Return 
        \EndIf
        \For{each $c \in \mathcal{C} \setminus \pi'$}
            \If{\texttt{verify($\pi' \cup \{c\}$,$(\mathcal{C},\mathcal{B},\mathcal{U})$)}}
               \State \Call{calculatePossibleOutcomes} {$\pi' \cup \{c\},(\mathcal{C},\mathcal{B},\mathcal{U})$}
            \EndIf
        \EndFor
    \EndProcedure
    \end{algorithmic}
   
\end{algorithm}


Our \texttt{calculatePossibleOutcomes} algorithm performs $O(|\mathcal{C}|)$ calls to \texttt{verify} which is $O(|\mathcal{C}|^2m)$, and the total number of calls is bound by the size of the permutation tree for $\mathcal{C}$ which is $O(|\mathcal{C!}|)$. Therefore, the entire algorithm is bounded by $O(|\mathcal{C}|^3 \cdot |\mathcal{C}|! \cdot m)$. However, the lower-bound is given by $\Omega(|\mathcal{C}|^2)$ providing improvements over the brute force approach as we verified experimentally in  Section~\ref{sec:mainresults}.

\subsection{Optimizations}

Function \texttt{verify} (Algorithm~\ref{alg:rcv_verify}) involves duplicate computations as an elimination order prefix $\pi'$ can be shared by multiple complete elimination orders $\pi$: for instance both $\pi_1=(A,B,C,D)$ and $\pi_2=(A,B,D,C)$ would eliminate the first two candidates $A$ and $B$ first and in the same order. Therefore, Loop 1 of Algorithm~\ref{alg:rcv_verify} involves repeated computations of the same elimination order prefixes. 

To avoid these redundant computations, we adapt Loop 1 of Algorithm~\ref{alg:rcv_verify} to use memoization to save election profile states for each partial elimination order $\pi'$. This allows us to improve the execution time of \texttt{calculatePossibleOutcomes} (Algorithm~\ref{alg:rcv_branch}) to run in  $O(|\mathcal{C}|^2 \cdot |\mathcal{C}|! \cdot m$) time. The lower bound is unchanged at $\Omega(|\mathcal{C}|^2)$.


An explanation of this bound is given next. Given a permutation tree of elimination orders for $|\mathcal{C}|$ candidates, it is clear at level $i$ the number of nodes is given by the permutation $\Perm{|\mathcal{C}|}{i}$. The height of the tree is also given by the number of candidates, $|\mathcal{C}|$. Therefore, we can compute the total number of nodes in a permutation tree with the summation below:
\[
\sum_{i=1}^{|\mathcal{C}|} \frac{|\mathcal{C}|!}{(|\mathcal{C}|-i)!} \leq |\mathcal{C}|! \cdot |\mathcal{C}| 
\]

Where $|\mathcal{C}|! \cdot |\mathcal{C}| $ is the number of nodes computation is performed on the worst case for Algorithm~\ref{alg:rcv_branch}. We assume $|\mathcal{C}| \geq 3$.

We can find the closed form of the summation:

\[
\sum_{i=1}^{|\mathcal{C}|} \frac{|\mathcal{C}|!}{(|\mathcal{C}|-i)!} = |\mathcal{C}|! \sum_{i=1}^{|\mathcal{C}|} \frac{1}{(|\mathcal{C}|-i)!}
\]

Using the Taylor series identity:

\[
    \sum_{i=0}^{\infty} \frac{x^i}{i!} = e^x
\]

We get

\[
    |\mathcal{C}|! \sum_{i=1}^{|\mathcal{C}|} \frac{1}{(|\mathcal{C}|-i)!} < |\mathcal{C}|! \cdot e
\]

And indeed for $|\mathcal{C}| \geq 3$:

\[
    |\mathcal{C}|! \sum_{i=1}^{|\mathcal{C}|} \frac{1}{(|\mathcal{C}|-i)!} < |\mathcal{C}|! \cdot e < |\mathcal{C}|! \cdot |\mathcal{C}|
\]

\subsection{Visualizing the outcome}

\begin{figure}[t]
    \centering
    \vspace{-1cm}
    \subfigure[Uncompressed Possible Elimination Order Tree \label{fig:raw_dag}]{\includegraphics [trim={200 200 200 200},width=0.24\textwidth]{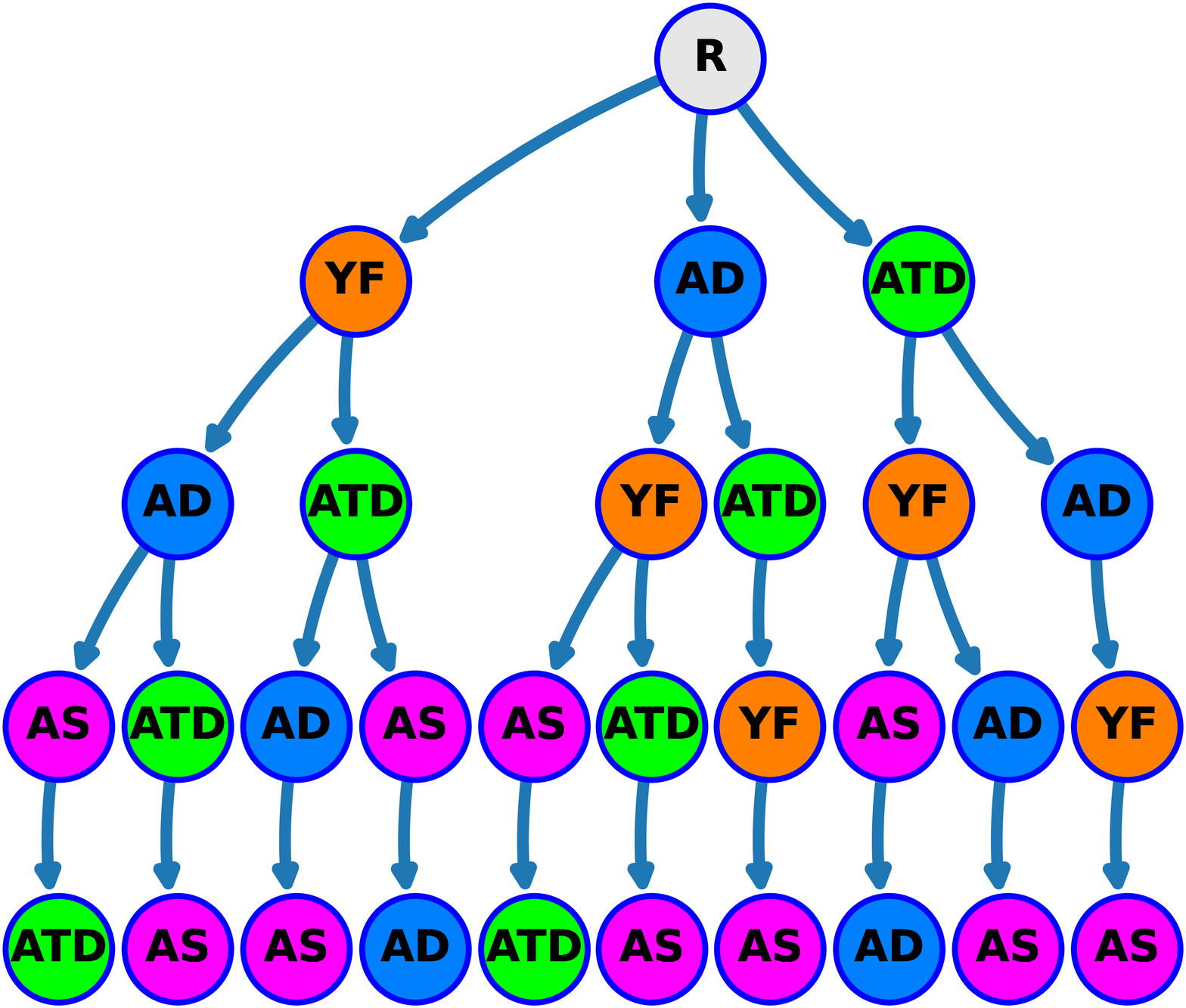}}
    \subfigure[Compressed  Possible Elimination Order DAG\label{fig:compressed_dag}]{\includegraphics [trim={200 200 200 200},width=0.24\textwidth]{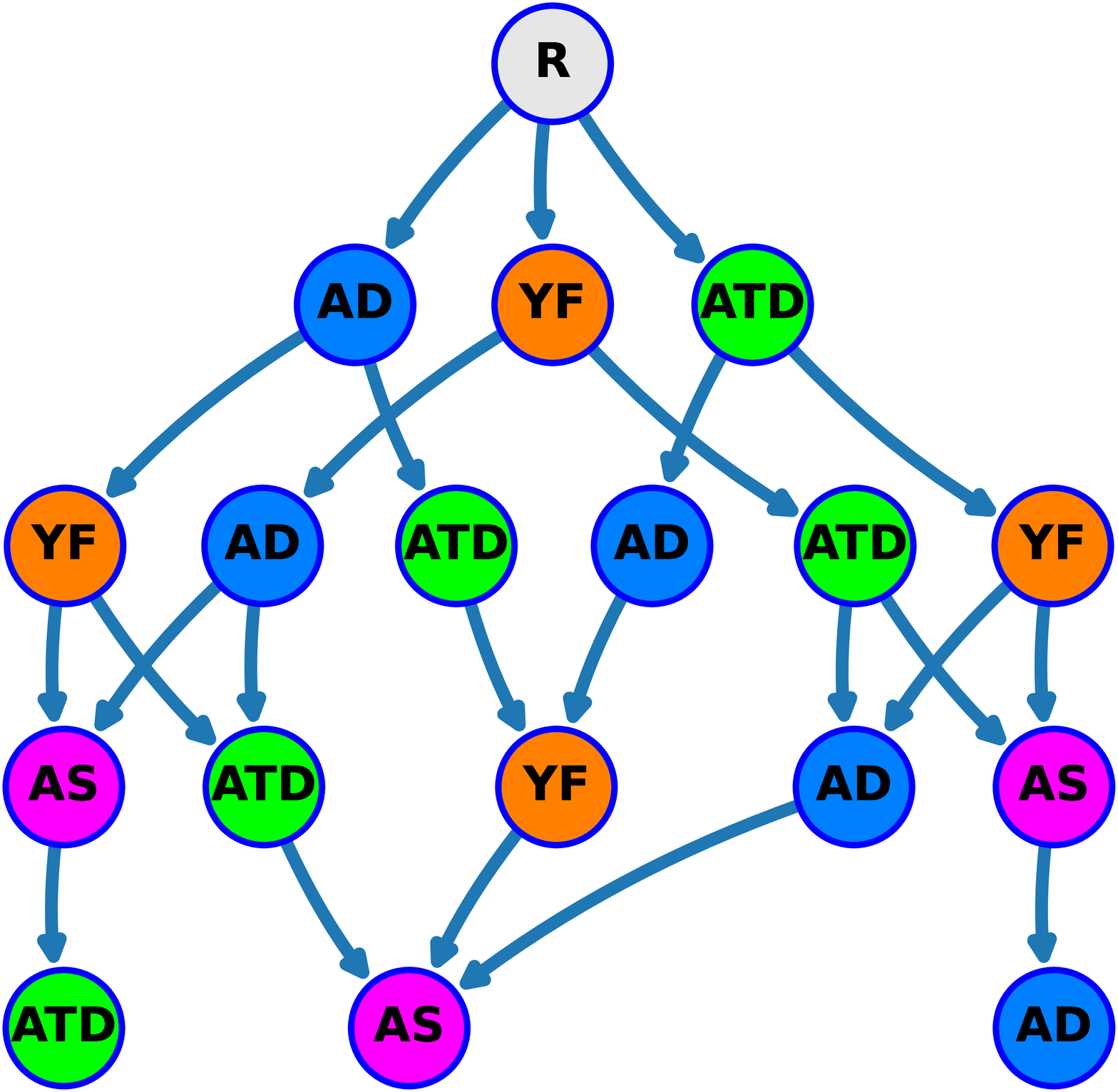}}
    \caption{DAG Compression for Possible Elimination Orders for the June 2021 New York Democratic Member of the City Council 16$^{th}$ Council District Primary Election }
    \label{fig:DAG}
\end{figure}

The resulting possible outcomes can be visualized as a tree, where each path from the root is a possible elimination path, and each leaf node is a possible winner. Figure \ref{fig:raw_dag} shows a possible elimination order tree for the June 2021 New York Democratic Member of the City Council 16$^{th}$ Council District Primary Election considering the information available the day after the election. Looking at the tree, it is obvious that the visualization could be made clearer by combining paths that lead to the same winners/final candidates. 

\cut{
We 
After obtaining $\mathcal{S}$ and creating a visualization tree, it is apparent that some subtrees structures are shared amongst other subtrees as shown in Figure \ref{fig:raw_dag}. For visualization purposes it is ideal to show that certain elimination orders converge to the same subtrees, hinting at inevitable outcomes with a certain number of missing ballots.
}

The problem of compressing trees has been studied extensively, notably in the context of XML trees. We adopted the XML compression method used in \cite{Buneman02_XMLPath} to our Possible Elimination Order trees for our visualization. More formally, given a tree $T$ we can iterate through each node $u \in T$ and obtain a string representation of the subtree $T'$ rooted at $u$. The string representation of $T'$ is then hashed with a common hash function (MD5), and inserted into a hash table. We then create a node for $u$ in a new graph $G$ that uses the hash of $T'$ as an identifier, and iterate through each child $v$ of $u$. This hashing process is performed for each $v$ if it is not already inserted in $G$, and an edge $(u,v)$ is then created in $G$. By allowing each node in $G$ to be identified by $T'$'s hash, we can ensure only one node for each $T' \in T$ is inserted into $G$. This method of compressing $T$ through repeated subtrees is commonly known as \emph{Directed Acyclic Graph (DAG) compression} \cite{Bille2015_treecompression}, and aims at creating the most minimal representation of tree $T$ in the form of a DAG. 

The resulting DAG is shown in \ref{fig:compressed_dag}. The simplified visualization makes it easy to identify that after the primary election day, and given the number of outstanding absentee ballots, the NYC City Council 16$^{th}$ Council District had 3 possible winners, and that candidate Althea Stevens (AS) was guaranteed to be one of the final two candidates.

\subsection{Refining the results}
\label{sec:postprocess}

Our \texttt{calculatePossibleOutcomes} algorithm identifies all possible winners of an election, but does not provide their relative chances of winning. Such a computation is complex: it requires considering all combinations of bindings for ballots in $\mathcal{U}$ including void and exhausted possibilities, a problem that is related to the knapsack problem. 

We propose a preliminary approach to provide some intuition of the relative chances of each candidate. Our approach involves a small modification of the \texttt{verify} function to output, for each elimination path, the number of unknown ballots that have to be bound to the winning candidate. For each candidate, we then calculate the minimum such value. This gives us an estimate of the minimum number of unknown ballots on which the candidate needs to lead to have a chance of winning the election.

The modification works by tallying all originally unbound ballots that contain votes for the winning candidate $c$ in a complete elimination order $\pi$ before \texttt{verify} returns. We ensure all missing ballots have at least one ranking.\footnote{This works by assigning one ranking to each empty unbound ballot such that the assignments to $c$ are minimized and the ordering of $\pi$ is preserved.} This value is then checked against previous tallied counts of originally unbound ballots that $c$ needed to win in other elimination orders, and the minimum value is recorded. We report on results and insights from this computation for some chosen NYC primary elections in  Section~\ref{sec:casestudies}.

Note, that our process assumes that all unknown ballots must be assigned to at least one candidate. As such we are not capturing all possible scenarios (e.g., if all, or some, absentee ballots are void or exhausted). We plan to investigate tighter bounds on the number of unknown ballots needed for a candidate $c$ to win in future work.

\cut{

We then return this value from \textttf{verify} and the caller of \textttf{verify} will update the minimum bound unknown ballots for the winner of $\pi$ if the new result is lower than the previous minimum.

We also propose an additional constraint on our unknown ballots - that is each unknown ballot must be assigned to at least one candidate before we can use the value returned from \texttt{verify}. This means we have an additional post-processing step where the empty ballots (if they exist) in $\mathcal{U'}$ are each assigned a single candidate before tallying the ballots in $\mathcal{U'}$ that contain marks for the winning candidate.
These empty ballots are assigned a new candidate by simply running $\mathcal{B'} \cup \mathcal{U'}$ through the standard RCV counting algorithm (\textttf{CountRankedVotes($\mathcal{C},\mathcal{B'} \cup \mathcal{U'}$)}) but with the modification that before each elimination round ends, we allocate the the maximum number of empty ballots from $\mathcal{U'}$ to the losing candidate such that this candidate still loses in the current round. We continue this process until all rounds are complete. In the case that more empty ballots remain after all rounds of this allocation complete, we divide the remaining empty ballots evenly across the candidates to preserve the elimination order. This ensures the winning candidate is assigned the absolute minimum amount of these empty ballots.
}
\section{Evaluation}

 We evaluated our proposed algorithm to compute all possible outcomes of a RCV election given partial ballot information by applying it to the contests in the 2021 New York City Primary elections. To validate the efficiency of our branch and bound mechanism, we also compared it to the brute-force approach. 
 
 In this section we report on our results, detailing the novel information that our partial vote counting algorithms allows to infer from the election profile available on election night. Our results (Section~\ref{sec:mainresults}) show that we would have been able to call winners in  21 of the 52 NYC primary contests that had more than two candidates as soon as election night. 
 In some cases our algorithm was able to identify results even when the election nights results were very close. For instance, in the NYC $40^{th}$ City Council District Primary, we were able to identify the winner even though she only had 25\% of first-choice votes on election night and two other candidates had 20\% each (see Section~\ref{sec:casestudies}). 

 We analysed the possible outcomes of several  June 2021 NYC primary  elections in Section~\ref{sec:casestudies} to 
 highlight the insights that our algorithm could have provided stakeholders: voters, candidates, political observers, news organizations, as early as election night, rather than them having to wait several weeks for the full results to be tallied.
 
\subsection{Dataset}

 We use the public election data from \cite{nycdata} for our experiments. The election data contains ballot-level data for 63 contests of the June 2021 NYC primaries. Eleven of these races had only two candidates; in such case the RCV vote count reverts to simple majority voting and our algorithm is not needed. In the rest of this paper, we provide results for the 52 June 2021 NYC primaries with more than two candidates.
 
 For each election, we consider the set of known, bound, ballots $\mathcal{B}$ to be the in-person ballots, the results of which were available on election night or the following day. In fact, the only publicly available results for more than a
 week after the election were a tally of first-choice rankings in $\mathcal{B}$ for each candidate.
 
 We consider the set of unknown, unbound, ballots $\mathcal{U}$ to be all absentee ballots, affidavit ballots and emergency ballots which had to be cured several days after the election. Voters can only vote in the primaries of one party, and the races on their ballots depend on their residence and party affiliation. We analysed each absentee, affidavit, and emergency ballot to identify which races it contains to have a correct number of unknown ballots for each race. 


\subsection{Experimental Setup}

Our algorithms are implemented in Python 3.9.5 and executed on an Intel Core i5-8250U CPU with 32.0 GB of RAM. Our implementation is single-threaded. For elimination graph visualizations, we use Python's NetworkX library with Graphviz. Our code is open-sourced and can be found at \url{https://github.com/ameliemarian/RCV}

We ran our tests for each election using the \texttt{calculatePossible\-Outcomes} algorithm and set a 2 hour timeout for each. When a timeout occured in the \texttt{calculate\-Pos\-sibleOutcomes} algorithm, we removed the first-round ranked candidates with below $5\%$ of the ballots from the election profile in order to run the contest with fewer candidates, and thus explore a much smaller search-space and re-run the comparison. For the contests where both algorithms finished within our timeout period, we compared the elimination graphs and minimum-bound absentee ballot counts for correctness.

\subsection{Possible election outcomes of the June 21 NYC Democratic Primaries}
\label{sec:mainresults}

\begin{table*}[t]
\scriptsize{
    \begin{tabular}{| l | l | l | p{3cm} | p{1.2cm} | r | p{0.8cm} | l |}
    \hline
    $\mathcal{|C|}$ & $\mathcal{|B|}$ & $\mathcal{|U|}$ & Runtime(s) & Runtime(s) & Percent & Possible  & Contest \\
    &&&\texttt{calculatePossibleOutcomes}&Brute Force & Speedup & Winners&\\
    &&&Algorithm& Algorithm& &\\\hline \hline

        3 & 17247 & 2174 & 0.013 & 0.016 & 23\% & 1 & Kings Dem City Council D45 \\ \hline
        3 & 711166 & 165689 & 1.551 & 1.565 & 1\% & 1 & New York Dem Public Advocate \\ \hline
        3 & 15069 & 3588 & 0.044 & 0.048 & 9\% & 2 & Bronx Dem City Council D12 \\ \hline
        3 & 13980 & 4350 & 0.041 & 0.043 & 5\% & 1 & Queens Dem City Council D31 \\ \hline
        3 & 164309 & 66172 & 1.302 & 1.193 & -8\% & 3 & Queens Dem Borough President \\ \hline
        3 & 11489 & 5206 & 0.062 & 0.079 & 27\% & 2 & Queens Dem City Council D28 \\ \hline
        4 & 14299 & 1371 & 0.032 & 0.051 & 59\% & 2 & Kings Dem City Council D42 \\ \hline
        4 & 15295 & 1464 & 0.063 & 0.074 & 18\% & 1 & Kings Dem City Council D34 \\ \hline
        4 & 10648 & 1594 & 0.027 & 0.051 & 89\% & 1 & New York Dem City Council D8 \\ \hline
        4 & 6824 & 1381 & 0.032 & 0.054 & 69\% & 2 & Kings Dem City Council D47 \\ \hline
        4 & 8963 & 2508 & 0.118 & 0.144 & 22\% & 3 & Bronx Dem City Council D16 \\ \hline
        4 & 11114 & 4884 & 0.252 & 0.312 & 24\% & 4 & Queens Dem City Council D24 \\ \hline
        4 & 17749 & 1872 & 0.025 & 0.044 & 76\% & 2 & Richmond Rep Borough President \\ \hline
        5 & 21486 & 1883 & 0.048 & 0.227 & 373\% & 2 & Kings Dem City Council D36 \\ \hline
        5 & 6147 & 1488 & 0.211 & 0.406 & 92\% & 1 & Queens Dem City Council D21 \\ \hline
        5 & 7577 & 1771 & 0.267 & 0.473 & 77\% & 4 & Kings Dem City Council D48 \\ \hline
        5 & 89407 & 22259 & 1.735 & 4.158 & 140\% & 3 & Bronx Dem Borough President \\ \hline
        5 & 8834 & 2504 & 0.224 & 0.515 & 130\% & 1 & Bronx Dem City Council D13 \\ \hline
        5 & 23770 & 6282 & 0.646 & 1.388 & 115\% & 1 & Richmond Dem Borough President \\ \hline
        5 & 7315 & 1267 & 0.095 & 0.23 & 142\% & 3 & Richmond Rep City Council D50 \\ \hline
        6 & 9663 & 1270 & 0.435 & 2.223 & 411\% & 1 & Kings Dem City Council D37 \\ \hline
        6 & 10254 & 1920 & 1.3 & 3.912 & 201\% & 1 & Kings Dem City Council D38 \\ \hline
        6 & 24113 & 5122 & 4.337 & 10.38 & 139\% & 1 & New York Dem City Council D3 \\ \hline
        6 & 8498 & 1937 & 1.023 & 3.328 & 225\% & 3 & Bronx Dem City Council D14 \\ \hline
        6 & 32370 & 7755 & 4.208 & 13.386 & 218\% & 1 & New York Dem City Council D6 \\ \hline
        6 & 14340 & 4030 & 2.307 & 6.924 & 200\% & 2 & Bronx Dem City Council D22 \\ \hline
        6 & 8819 & 3004 & 1.869 & 6.358 & 240\% & 2 & Queens Dem City Council D32 \\ \hline
        6 & 10457 & 5749 & 5.634 & 13.315 & 136\% & 6 & Queens Dem City Council D19 \\ \hline
        7 & 32087 & 3089 & 4.073 & 37.952 & 832\% & 2 & Kings Dem City Council D35 \\ \hline
        7 & 32969 & 3409 & 1.842 & 34.2 & 1757\% & 1 & Kings Dem City Council D39 \\ \hline
        7 & 202319 & 39102 & 87.762 & 410.548 & 368\% & 2 & New York Dem Borough President \\ \hline
        7 & 20688 & 5501 & 21.049 & 83.118 & 295\% & 4 & New York Dem City Council D5 \\ \hline
        7 & 14240 & 5848 & 18.715 & 74.606 & 299\% & 4 & Bronx Dem City Council D11 \\ \hline
        7 & 13369 & 6654 & 50.211 & 120.295 & 140\% & 7 & Queens Dem City Council D23 \\ \hline
        8 & 17898 & 1839 & 12.034 & 238.567 & 1882\% & 1 & New York Dem City Council D10 \\ \hline
        8 & 16769 & 1831 & 8.497 & 220.029 & 2490\% & 1 & Kings Dem City Council D46 \\ \hline
        8 & 12518 & 1552 & 7.553 & 178.915 & 2269\% & 2 & Bronx Dem City Council D18 \\ \hline
        8 & 25655 & 3948 & 59.435 & 584.203 & 883\% & 1 & Kings Dem City Council D33 \\ \hline
        8 & 7350 & 1737 & 23.033 & 215.668 & 836\% & 1 & Bronx Dem City Council D15 \\ \hline
        8 & 8394 & 4680 & 622.514 & 1341.583 & 116\% & 8 & Queens Dem City Council D20 \\ \hline
        8 & 11259 & 7498 & 1046.66 & 1948.612 & 86\% & 8 & Queens Dem City Council D25 \\ \hline
        9 & 19111 & 3046 & 101.881 & 3299.032 & 3138\% & 1 & New York Dem City Council D1 \\ \hline
        9 & 12026 & 3202 & 1055.637 & 6219.858 & 489\% & 3 & Richmond Dem City Council D49 \\ \hline
        9 & 13848 & 6886 & 5422.558 & [Timeout] & N/A & 9 & Queens Dem City Council D29 \\ \hline
        10$^{\S 3}$ & 756047 & 165688 & 199.23 & 1318.241 & 562\% & 3 & New York Dem Comptroller \\ \hline
        11 & 20280 & 2256 & 1701.439 & [Timeout] & N/A & 1 & Kings Dem City Council D40 \\ \hline
        12$^{\S 4}$ & 260695 & 31872 & 83.119 & 1951.516 & 2248\% & 2 & Kings Dem Borough President \\ \hline
        12$^{\S 5}$ & 20173 & 3373 & 12.619 & 48.691 & 286\% & 1 & New York Dem City Council D7 \\ \hline
        12$^{\S 4}$ & 17964 & 5106 & 261.314 & 912.354 & 249\% & 1 & Queens Dem City Council D27 \\ \hline
        13$^{\S 4}$ & 23193 & 2624 & 110.895 & 3064.412 & 2663\% & 3 & New York Dem City Council D9 \\ \hline
        13$^{\S 5}$ & 822410 & 165689 & 443.818 & [Timeout] & N/A & 3 & New York Dem Mayor \\ \hline
        15$^{\S 6}$ & 16097 & 3846 & 796.932 & 5281.628 & 563\% & 3 & Queens Dem City Council D26 \\ \hline
    \multicolumn{7}{c}{}\\
    \end{tabular} 
}
    \caption{Results of the \texttt{calculatePossibleOutcomes} and brute-force algorithms on the Election-night data of the June 2021 New York City Democratic Primaries}
    \label{tbl:results}
    \vspace{-0.5cm}
\end{table*}

Table~\ref{tbl:results} contains bound and missing ballot counts, runtimes, and counts of possible winners for each contest  with more than 2 candidates in the 2021 New York City Primaries. Our algorithm was able to identify the winners for 21 of the 52 contests using election night data. For 9 of these contests, these results were not a surprise, as the only possible winner was a candidate with over $50\%$ of the votes on election night and with not enough missing ballots to change the result. For another 12 of these contests, our algorithm correctly identifies 1 possible winner even though that candidate did not have over $50\%$ of the ballots in the first round on election night. In most cases we are able to reduce the number of possible winners to 2 or 3, although this was not always possible when the percentage of unknown ballots is large or the first round rankings were very close. 

\begin{figure}[t]
\centering
    \includegraphics [width=0.5\textwidth]{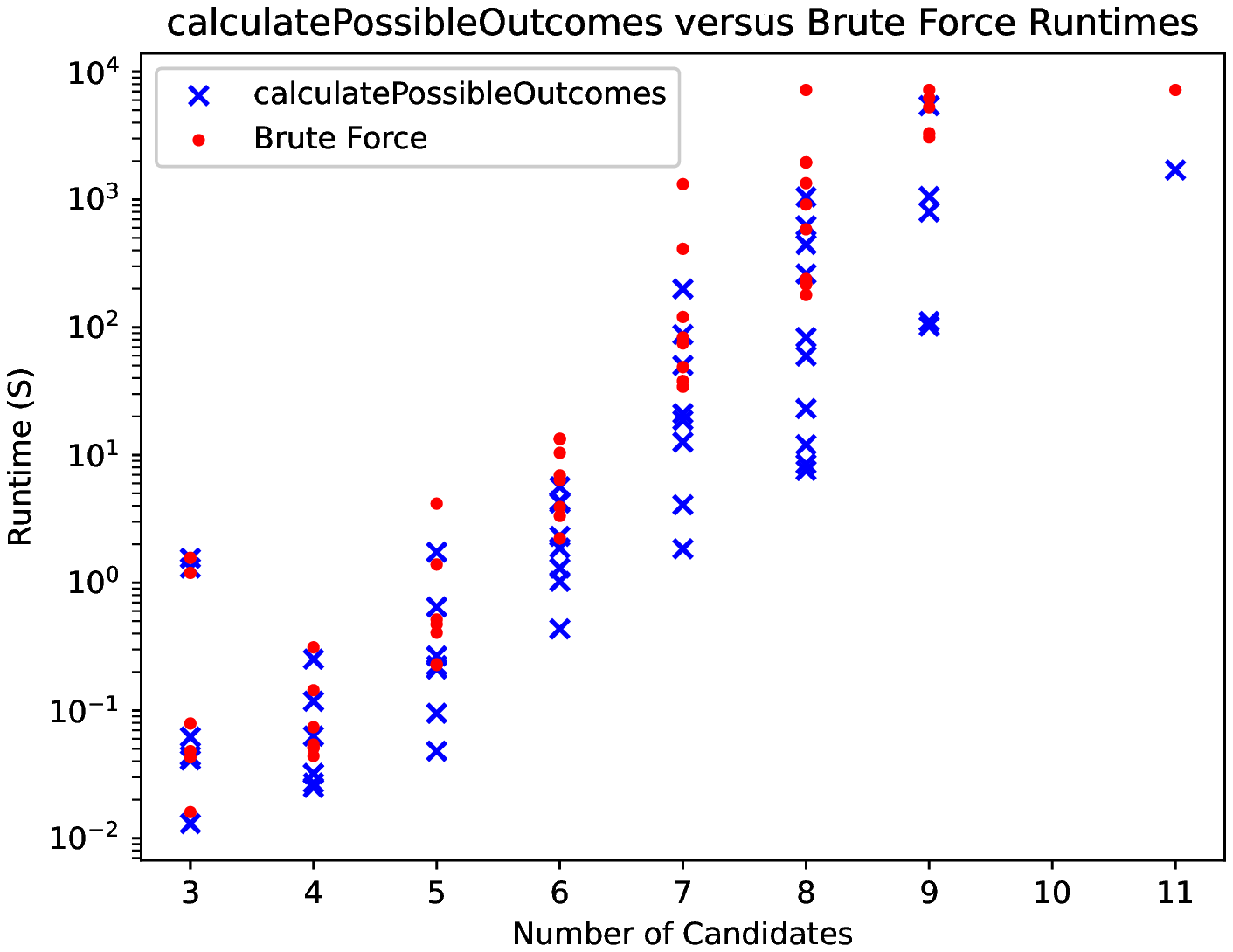}
  \caption{Runtimes (logarithmic scale) by numer of candidates  $\mathcal{|C|}$ of  \texttt{calculatePossibleOutcomes} and Brute Force Algorithm for all election races from Table~\ref{tbl:results}.} 
  \label{fig:runtimes}
    
\end{figure}

The runtime of our algorithm, shown in Figure~\ref{fig:runtimes}, is often significantly faster than that of the brute-force approach. For all races with fewer than 8 candidates \texttt{calculatePossibleOutcomes}  runs in less than 2 minutes (less than 10 seconds in most cases) and has an average speedup of 180\% compared to the brute-force algorithm. For races with 8 candidates and over, \texttt{calculatePossibleOutcomes} has an average speedup of over 1000\% compared to the brute-force algorithm. As the number of candidates grows, the number of elimination orders to consider grows exponentially. For elections that timed out within our 2 hour limit, we reduced the search space by pruning the number of candidates. We removed all candidates with less than 5\% of the vote as these were very unlikely to have a path to victory.\footnote{While there may be theoretical cases where a candidate with less than 5\% of first choice votes can win a RCV election, this has not happened in practice\cite{fairvote}} The number of candidates removed, $n$, for an election that timed out with $\mathcal{|C|}$ candidates is denoted in Table~\ref{tbl:results} as $\mathcal{|C|}^{\S n}$. In Figure~\ref{fig:runtimes} we present these elections as having $\mathcal{|C|}- n$ candidates and assigned a value of 2 hours to the timed out executions.

\begin{table}
\scriptsize{
        \begin{tabular}{| l | l | p{1.2cm} | p{1.5cm} | p{1.8cm} | p{1.3cm} | p{1.0cm} | l |}
        \hline
        Contest & Minimum Bound Absentee Ballots \\ 
        \hline
        New York Dem Mayor & \makecell{Eric L. Adams: 0 \\ Kathryn A. Garcia: 15776 \\ Maya D. Wiley: 66440} \\
        \hline
        Kings Dem City Council D36 & \makecell{Chi A. Osse: 0 \\ Tahirah A. Moore: 1726} \\
        \hline
        Kings Dem City Council D35 & \makecell{Crystal Hudson: 0 \\ Michael Hollingsworth: 2121} \\
        \hline
        Kings Dem Borough President & \makecell{Antonio Reynoso: 0 \\ Jo Anne Simon: 18564} \\
        \hline

        New York Dem Comptroller & \makecell{Brad Lander: 0 \\ Corey D. Johnson: 20589 \\ Michelle Caruso-Cabrera: 125518} \\
        \hline

        \end{tabular}
        }
        \caption{Minimum number of  bound absentee ballots needed to win for each possible winner across selected contests of the June 2021 New York City Democratic  Primaries}
        \label{tbl:min_bound}
    \end{table}

\subsection{Case Studies}
\label{sec:casestudies}

In this section, we highlight some interesting results and insights we were able to infer by applying our algorithm on the 2021 NYC Democratic Primary election-night data.

In Table~\ref{tbl:min_bound}, we provide some of the results of the postprocessing step of Section~\ref{sec:postprocess}. These results show, for selected races, the minimum number of unknown (absentee) ballots that each candidate needs to win to have a path to victory. Note that this does not guarantee that a candidate's victory if they receive that number of absentee votes; it gives a lower bound of the number of votes they would need to win in the most favorable possible elimination path. This explains why some candidates have a minimum number of 0 absentee ballots. From this table, and the total number of absentee ballots available in Table~\ref{tbl:results}, it seems obvious that while some candidates have a mathematically possible path to victory, the chances of them winning are slim. For example, in the Comptroller race, Michelle Caruso-Cabrera would need to win at least 75\% of absentee ballots, an unlikely outcome as she only won 13.5\% of the in-person ballots (Table~\ref{tbl:min_bound}).

\subsubsection{June 2021 NYC Mayoral Democratic Primary}

The results of the Mayoral Primary were understandably the most awaited results of the primary. That particular race turned out to be a perfect illustration of the benefits and drawbacks of RCV. On election night, Eric Adams was leading with 31.66\% of ballots, Maya Wiley was second with 22.22\% and Kathryn Garcia third with 19.58\%. Andrew Yang was a distant fourth with 11.66\%. The data made public on election night  was limited to first choice votes. It was clear that vote transfers would decide the election result and could lead to surprises. A week after the election, a count of RCV was performed {\em only on in-person votes}. The outcome showed that once Andrew Yang was eliminated in the third-to-last round, his vote transfers were enough for Kathryn Garcia to edge out Maya Wiley {\em by only 400 votes}, only to lose to Eric Adams in the last round. 

These partial results raised more questions than they answered: over 165,000 absentee ballots were still to be counted. What if Maya Wiley were to be in the final round against Eric Adams? Would she win against him? How would votes transfer in other possible election orders? Could the final two be Maya Wiley and Kathryn Garcia? What would happen in that scenario?

\begin{figure}[t]
\centering
    \includegraphics [trim={900 1100 800 1200},width=0.5\textwidth]{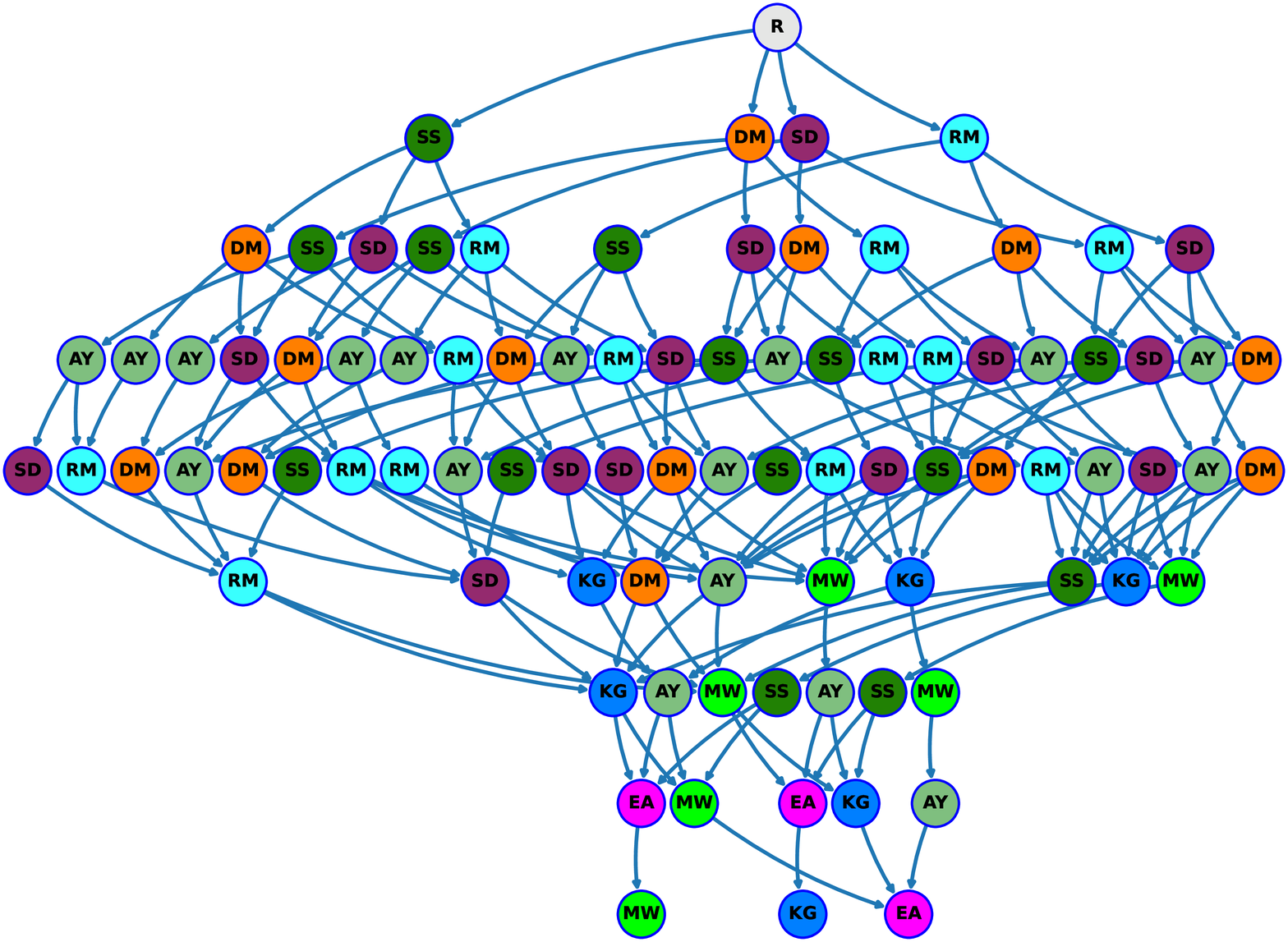}
  \caption{Possible Elimination Orders for the 2021 NYC Democratic Mayoral Primary }
  \label{fig:NYC_mayoral}
    
\end{figure}

Our algorithm would have been able to answer some of these questions: Figure~\ref{fig:NYC_mayoral} shows the possible elimination orders for the mayoral primary. All three candidates, Eric Adams, Kathryn Garcia, and Maya Wiley had paths to victory, but Eric Adams was guaranteed to finish $1^{st}$ or $2^{nd}$.

Table~\ref{tbl:min_bound} provides more insights: our algorithm identifies Maya Wiley needing a minimum of 66,440 ($40\%$) of the absentee ballots for a path to victory while Kathryn Garcia would have needed a minimum 15,776 ($9.5\%$) of the absentee ballots for a path to victory. Eric Adams had a path to victory that did not require him to win any absentee ballots (for instance an unlikely scenario where all absentee ballots are shared among minor candidates). This would have shown that all three were potential winners, but Maya Wiley had to appear before the other two candidates in a much larger number of absentee ballots to have a chance at winning.

\subsubsection{June 2021 Kings Democratic Member of the City Council $36^{th}$ Council District Primary}

On election night, candidate Tahirah A. Moore was tied in second place with Henry L. Butler (each at 4720 and 4721 ballots respectively) in the first round. The first place candidate (and eventual winner) Chi A. Osse was ahead of them by 2969 ballots. Our algorithm identifies Tahirah A. Moore as the only other possible winner, despite Henry L. Butler having the same number of votes,  when 1883 absentee ballots were present (approximately $8\%$ of all the cast ballots). Moore needed a minimum of 1726 absentee ballots  for a possible path to victory. Therefore, on election night our algorithm could practically identify the winner in this contest as it is unlikely Moore would appear above Osse in $92\%$ of the absentee ballots.

\subsubsection{June 2021 Kings Democratic Member of the City Council $35^{th}$ Council District Primary}

On election night, candidate Michael Hollingsworth was in second place, closely trailing leading candidate (and eventual winner) Crystal Hudson  by 1291 ballots in the first round. Our algorithm identifies Hollingsworth as the only other possible winner,  needing a minimum of 2121 absentee ballots for a possible path to victory. However, there exists only 3089 absentee ballots; it seems unlikely that Hollingsworth would receive over $68\%$ of the missing ballots and have a path to victory. Therefore, by election night, we could have inferred that there was likely only 1 winner for this contest even though Hollingsworth and Hudson had $34.45\%$ and $38.49\%$ of the election night ballots respectively.

\subsubsection{June 2021 Kings Democratic Borough President Primary}

On election night, candidate Robert E. Cornegy Jr. was in second place, trailing behind leading candidate (and eventual winner) Antonio Reynoso by 22955 ballots (9\% of the total known ballots) in the first round. In third place was Jo Anne Simon, who was 3927 votes behind Cornegy. Our algorithm identifies Simon as the only other possible winner to the eventual winner (Reynoso), needing a minimum of 18564 absentee ballots for a possible path to victory. These results show that is was impossible for Cornegy to win, since he had no possible path to victory, despite being in second place on election night. However, Simon who was in third place only needed about $58\%$ of the absentee ballots for a path to victory.



\subsubsection{June 2021 Kings Democratic Member of the City Council $40^{th}$ Council District Primary}

On election night, candidate (and eventual winner) Rita C. Joseph was in first place with 5060 ballots ($25.23\%$ of the total known ballots) in the first round. In second and third place are candidates Josue Pierre and Kenya Handy-Hilliard, each with 4073 and 3849 ballots respectively. With 2275 absentee ballots, it might seem that both of these candidates are likely to have a path to victory. However {\em our algorithm identified Joseph as the only possible winner on election night} (Table~\ref{tbl:results}).

\cut{
\begin{figure}[H]
    \centering
    \centerline{
    \includegraphics [width=1.5\linewidth]{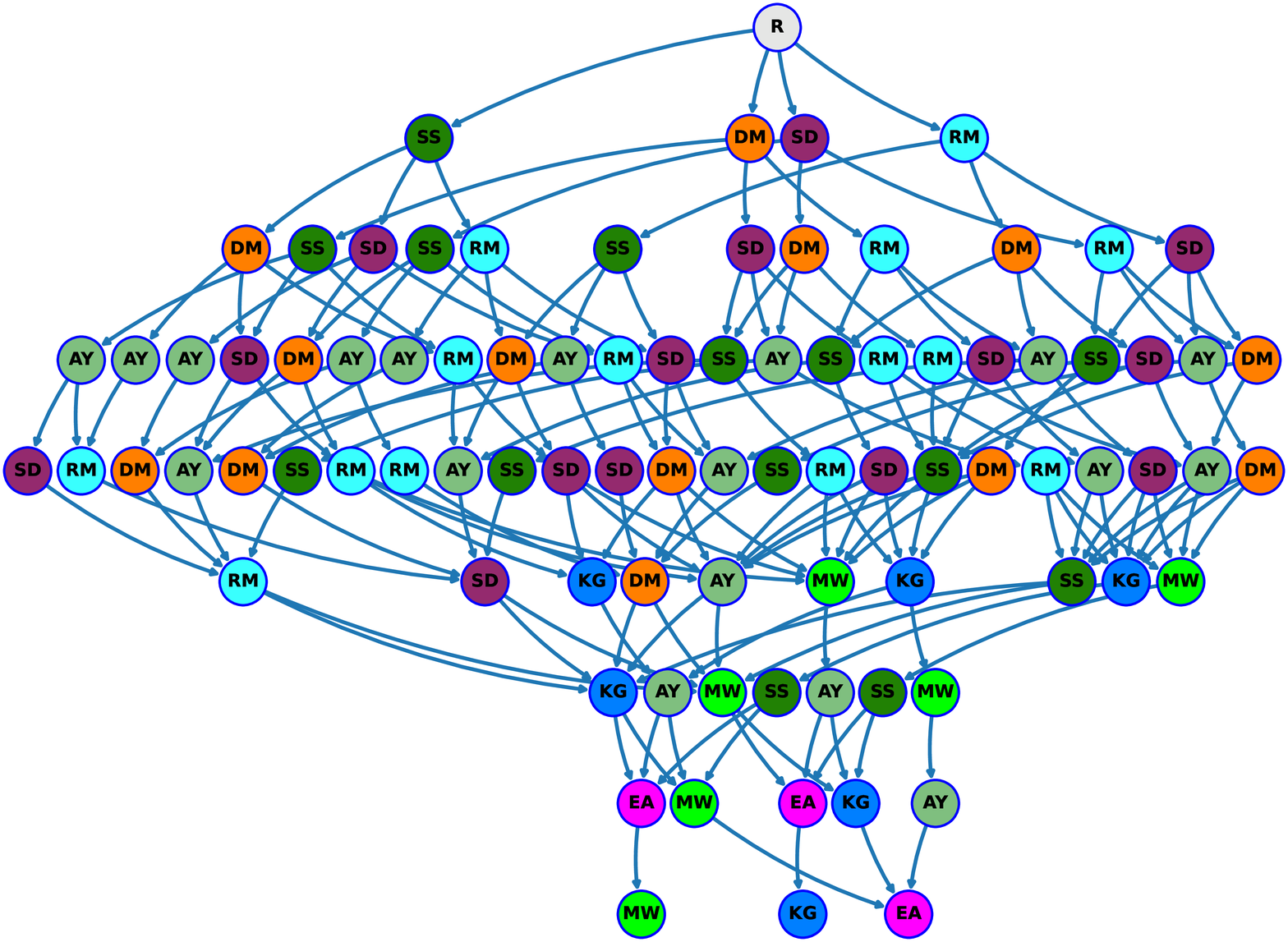}
    }
    
  \caption{2021 NYC Democratic Mayoral Elections }
  \label{fig:NYC_mayoral}
    
\end{figure}
}

\cut{
\subsection{Summary of evaluation}
Our results show that there is a clear benefit to counting partial results of RCV elections even before all ballots are tallied, as the results of many races can be identified even with outstanding ballots. For other races, while some candidates can mathematically win the election, our algorithm gives insight as to the likelihood of such a scenario.

This information should prompt localities that use RCV as their election mechanism to identify all possible winners when reporting preliminary results and to make cast ballot data available to the public as soon as possible for transparency. While some districts share cast ballot record data as soon as they are tallied (e.g., San Francisco), others only make that information public long after the election (NYC released the data two months later), or do not release it at all. To ensure fair and trusted RCV election, election districts should commit to transparent and timely releases of RCV ballots.  }
 \section{Conclusion}

We presented an algorithm to identify possible winners of Ranked-Choice Voting elections when all ballots have not yet been tallied. Our approach allows for stakeholders to get insights on the possible outcome of the election earlier than in the current real-world scenarios where either all ballots need to be tallied before the RCV round-by-round process is performed, or a preliminary RCV tally is performed on  ballots known on election night, only considering one possible election outcome and potentially producing misleading information. 

The techniques proposed in this paper identify all still-possible winners of a RCV election but do not assign probabilities to their chances of winning. An interesting extension of our work would be to integrate probabilistic query processing techniques~\cite{suciu2011probabilistic} to RCV vote counting to provide better insights on the election results.  

Our work provides critical tools to provide clarity to voters and candidates on the election outcome and to increase transparency and trust in RCV election processes. This should prompt localities that use RCV as their election mechanism to identify all possible winners when reporting preliminary results and to make cast ballot data available to the public as soon as possible for transparency.


\bibliographystyle{ACM-Reference-Format}
\bibliography{main}

\end{document}